%% file: ms.tex
\documentclass{pbbasic}
\pdfoutput=1
\usepackage{dcolumn}
\usepackage{multirow}

\newcommand{\zm}{\operatorname{sim}}

\newcommand{\brname}[2]{\mbox{\sf #1v#2}}
\newcommand{\brange}[4]{\brname{#1}{#2} = [ #3 , \ldots , #4]}

\begin{document}

%---------------------------------------------------------
%
%                         FRONT
%

\title{Computer activity learning from \\ system call time series}
\author{Curt Hastings and Ronnie Mainieri \\ [0.5ex] {\small \texttt{\{cah,rxm\} at permissionbit.com}} \\ [0.5ex] {\small PermissionBit, 1750 Tysons Blvd.~Ste.~1500, McLean VA 22102}}
\date{26 October 2017}

\maketitle

\begin{abstract}
    \input{abstract}  
\end{abstract}

%---------------------------------------------------------
%
%                       MAIN BODY
%

% intro
\input{intro}

\input{detection}

\input{performance}

\input{discussion}

%---------------------------------------------------------
%
%                       BACK MATTER
%

\urlstyle{same}
\sloppy
\printbibliography
\fussy

\end{document}

%% file: abstract.tex
Using a previously introduced similarity function for the stream of system calls generated by a computer, we engineer a program-in-execution classifier using deep learning methods.  Tested on malware classification, it significantly outperforms current state of the art.  We provide a series of performance measures and tests to demonstrate the capabilities, including measurements from production use.  We show how the system scales linearly with the number of endpoints.  With the system we estimate the total number of malware families created over the last 10 years as \num{3450}, in line with reasonable economic constraints.  The more limited rate for new malware families than previously acknowledged implies that machine learning malware classifiers risk being tested on their training set; we achieve $F_1 = 0.995$  in a test carefully designed to mitigate this risk.

%% file: intro.tex
% intro.tex

\section{Introduction}

Current practices in cybersecurity do not scale, reflecting an asymmetry in the ability to automate that favors the attacker. There is an emerging consensus in the industry that artificial intelligence (AI) could recover the advantage, but even with advances in other domains performant AI systems for cybersecurity remain elusive despite concerted efforts to develop them.

We previously described a method for embedding system call traces in a metric space using techniques that respect constraints to deployment \cite{dots}. Since Forrest's 1994 report \cite{forrest1994self} there has been a consensus among academics that sequences of system calls are the preferred data source for characterizing the behavior of a computer \cite{forrest1996sense, canali2012quantitative, blokhin2013malware, stokes15, kolosnjaji2016deep, athiwaratkun2017malware}, yet we are not aware of any viable systems that are based on this technique. We believe this is because until recently the data were difficult to obtain and the necessary statistical techniques did not exist. Here we engineer an in-execution malware detection system that significantly outperforms current state of the art, and we demonstrate that it will scale linearly to networks with a few million endpoints.

We first demonstrate the properties of our similarity function and show that we can identify relationships between versions of common programs that differ by several years of development. The relationship between resource expenditures and security was explored in Van Dijk \emph{et al.} \cite{flipit}; an in-execution system for detecting program similarity would eliminate much current reverse engineering cost (and would be competitive with existing systems). It would also force attackers to write their tools anew for each campaign. The observed reliance of attackers on a small number of malware code bases suggests that current tools used by defenders are unable to impose this cost.

We next characterize the distribution of expected activities that will be observed using our system under real usage conditions. We show that no more than a few thousand non-malicious activities are expected over long time periods (or across wide networks). The rate at which new malware programs appear is also much lower than generally asserted. Using the system, we conducted over 2.6 million experiments with a well-sampled corpus of live malware. We estimate the total number of malware families created for the Windows platform over the last 10 years to be \num{3450}, in line with reasonable economic constraints to software development; of these we have identified \num{1111}.

The broad coverage of the set of both malicious and non-malicious programs permits us to use features derived from our embedding as input to bootstrap a simple deep belief network that classifies programs that are dissimilar to any in our malware and non-malware corpora. The small number of malicious program families implies that care must be taken when testing the performance of AI subsystems; with millions of samples even rare programs have with high probability many homologs in the training set.  We carefully test our system for its ability to generalize by using malware sampled from the wild several months after we generated the data that was used to train the machine learning subsystems and by isolating the performance of the deep belief net by considering only the performance on samples dissimilar to any in our list of known malware.

Finally, we extrapolate the performance of our system to very large networks. We show that the similarity function and deep belief network together create an AI system that can analyze terabytes of system call sequence data per week. As we are interested in creating a system for live indentification of malicious activity on endpoints (often described as threat hunting in the trade), our work differs from others that boost performance by relying on data only available in an instrumented laboratory environment or that cannot be collected when the majority of devices are mobile. While these assumptions simplify development, the importance of endpoint context in determining the behavior of a program implies that designing for the more stringent requirements associated with in-execution monitoring is worthwhile. Our results demonstrate that a performant and scalable AI system can be constructed without emulation, hooking, VM introspection, or sandboxing at the network edge.

% byte rate
% (endpoint count) * (50 hours/week) * (60 min/hr) * (4 uploads/min) * (10 kb/upload) * (100:1 compression) = 10 TB

%% file: detection.tex
\section{Detection system}

In our system, each endpoint runs a \emph{sensor}, a data collection program, that transmits data to the \emph{compute server}, an analysis daemon that identifies traces of any executing malware.  The compute server has access to several databases and machine learning subsystems that it uses to organize the data it receives into detections. The detections are shipped to a security information and event management (SIEM) system for visualization and decision-making.  Each detection is assigned a probability that it corresponds to malicious execution on the originating endpoint. 

System calls are collected in \emph{batches} by a sensor on the endpoint using facilities provided by the operating system (the ETW framework in the case of Windows).  The sensor collects all calls issued in a 5 second interval, corresponding to about \num{400000} calls.  Within a batch, the streams of calls from each thread are grouped into sequences of 5-grams (referred to as \emph{words}).  The sensor then waits for a short period of random duration (which could be zero) before collecting the next batch of system calls.

Observing a set of computers under use for many hours under diverse usage conditions shows that the distribution of words follows an apparently long-tailed distribution.  The observation frequency of a word is proportional to a function $h$ that goes to zero with ever larger rank $r$ but at a rate slower than the tail of a normal distribution.  The expectation arising from the behavior of complex systems is that $h(r) \sim r^{-\alpha}$ for some $\alpha \ge 1$, but we have instead observed that in professionally maintained networks the rank distribution $h(r) \sim e^{-\alpha r}$ for $\alpha$ positive (although the sum of two exponentials provides a better fit to the more frequent words).  It is this fast decay that makes it possible to use a small collection of words as the basis in which to collect information from a computer.  This was an unexpected result.  

The words in each batch are processed on the endpoint into more informa\-tion-rich and compact structures that are better suited for additional processing by machine learning algorithms.  All words belonging to the same process in a batch are summarized into a pair of vectors: one, labeled zeta, is a 141-dimensional vector of reals $\zeta$ that reflects the information from the most frequent words; the other, labeled mu, is a bag of word-counts $\mu$ that reflects the information from the less frequent words.  Together the two form the feature vector $\left( \zeta, \mu \right)$ that is used in further processing.  The feature vector is typically stored with the timestamp and some metadata (process names and their numerical identifiers, for example) that may be useful in forensic work and in a SIEM user interface.  The construction of the feature vectors has been described in detail by \textcite{dots}, where they are called process dots. 

\subsection{Similarities}

The similarity $\zm(\cdot \, , \cdot)$ between two feature vectors $p_a = (\zeta_a, \mu_a)$ and $p_b = (\zeta_b, \mu_b)$ can be defined as a linear combination of the similarities between the zeta and the mu parts.  Because of the probabilistic construction of both zeta and mu, small corrections are made for edge cases of extreme values as previously described \cite{dots}, but essentially each similarity is cosine-like, computed from the scalar product of the components suitably normalized:
\begin{equation}
    \zm(p_a, p_b) =
    \theta_\zeta
    \langle \zeta_a, \zeta_b \rangle
    +
    \theta_\mu
    \langle \mu_a, \mu_b \rangle
    \,.
    \label{eqSim}
\end{equation}
The real numbers $\theta_\zeta$ and $\theta_\mu$ are chosen so that each dot product contributes approximately equal parts to the total similarity.  As zeta and mu belong to different vector spaces, their scalar products are different operations.

The compute server uses the similarity function $\zm(\cdot, \cdot)$ to aggregate some of the feature vectors from a few batches into \emph{tight clusters}.  These are clusters in which most of the pairs of vectors have similarity above a threshold $\theta_S$ (and a few additional graph-theoretical properties are present \cite{dots}).   Tight clusters can be compared by comparing the feature vectors within them.  Given two clusters $c_a$ and $c_b$, their similarity $\zm_{TC}$ can be defined as the average similarity among their constituent feature vectors, with $p_i \in c_a$ and $p_k \in c_b$:
\begin{equation}
    \zm_{TC}(c_a, c_b)
    =
    \frac{1}{|c_a| |c_b|} \sum_{p_i, p_j} \zm(p_i, p_j)
    \label{eqTCsim}
\end{equation}
where the sum runs over all possible pairs of feature vectors in the tight clusters and $|c_\alpha|$ is the number of feature vectors in cluster $\alpha$.  The actual implementation of the sum needs to correct for extreme values and can be constructed as an estimate computed in constant time.  

\begin{figure}
    \includegraphics[width=\textwidth]{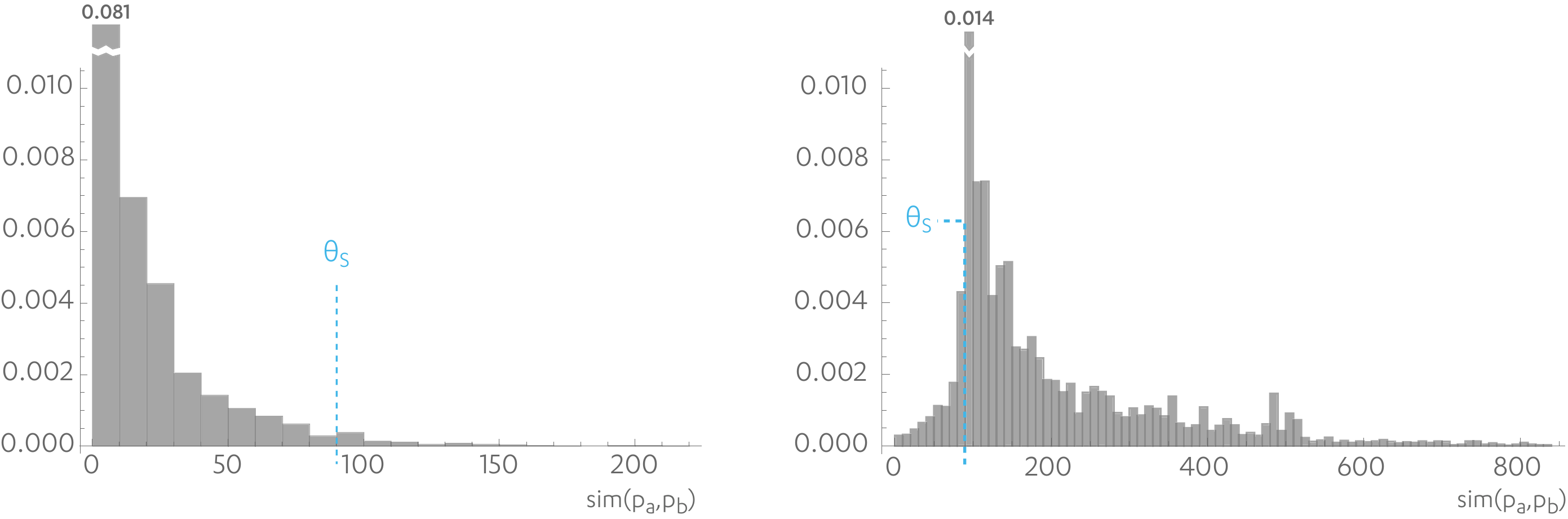}
    \caption{Probability distribution of similarities among feature vectors from different tight clusters.  Left, 250 feature vectors were sampled from a small collection of 45 tight clusters, and similarities were calculated over all pairs consisting of two vectors from different clusters.  Right, the same set of tight clusters are used to show the distribution of similarities for pairs of feature vectors within a tight cluster. In the distribution on the right, 23\% of the similarities are smaller than the $\theta_S$ threshold.}
    \label{figDist01}
\end{figure}

The similarity function $\zm(\cdot \, , \cdot)$ when used with tight clusters  already provides a performant malware detection system.  \autoref{figDist01} shows the probability distribution for the distance among feature vectors drawn from different clusters.  The majority of the feature vectors are dissimilar (have zero similarity), with the distribution rapidly falling off. In  \autoref{figDist01}, left, only 0.85\% of the similarities are larger than 90 (the threshold used to label two feature vectors as similar).  Assume we could establish two lists of tight clusters: $G$, formed by feature vectors produced by non-malicious programs ${\cal G} = \{ g_1, g_2, \ldots \}$ and $B$, formed by feature vectors only ever produced by malicious programs ${\cal B} = \{b_1, b_2, \ldots \}$.  If any of the programs in ${\cal B} \cup {\cal G}$ execute and produce a series of feature vectors yielding a tight cluster $c$, we can check from which set it originates by using the largest similarity to select a set.  If 
\begin{equation}
    \max_{b \in \cal B} \zm_{TC}(c, b) 
    >
    \max_{g \in \cal G} \zm_{TC}(c, g) 
    \label{eqDecision}
\end{equation}
then $c$ originates from malware, and the endpoint that produced it is likely compromised.

If the distribution in \autoref{figDist01}, left, were a good estimate for the distribution of similarities between feature vectors in different tight clusters (taken over tight clusters from all processes), then the fraction of tight clusters that would be incorrectly considered similar by the decision procedure in \autoref{eqDecision} would be 0.85\% (at $\theta_S = 90$).  Such a malware detection procedure would be immune to obfuscations of executable files, and it would be competitive with existing detection systems.  The distribution in \autoref{figDist01}, right, also shows the need for tight clusters.  If we applied the decision procedure in \autoref{eqDecision} to the feature vectors without first grouping them into clusters, around 23\% of the decisions would be misclassifications.

\subsection{Program families}
\label{secBrowserEx}

We next provide an example that illustrates the use of feature vectors to distinguish between evolving versions of two open source code bases in the same program class. The programs are Chrome \cite{openhubchrome} and Firefox \cite{openhubff}, each of which has about 17 million lines of code (loc, used to estimate the size of a code base; 17 Mloc is about \num{5000} developer-years of effort) as of this report. The different releases of each browser are each treated as a separate program identified by their version number, as in Chrome v12 or Firefox v3, for a total of 11 programs. We profiled Chrome versions 12, 17, 21, and 35 and Firefox versions 3, 17, 20, 28, 32, 39, and 44 using executables that did not require installation \cite{portableapps}. We refer to any of these 11 programs as a browser program.

Each browser program was exercised by visiting a variety of websites.   The process was repeated six times for each version.  During the experiment, stock Windows processes may have also been executing, as no effort was made to limit them.  Batches of system calls were collected (5 second batches with Poisson distributed wait between them with mean 3 seconds) and uploaded, but only data from the last few dozen were used to generate the matrix.  When Chrome is launched, multiple processes are created as part of its sandboxing functionality.  The feature vectors were collected for all processes that were labeled \texttt{chrome}.  To avoid the confusing visual banding this may create, the feature vectors from Chrome were sorted first by their version number and then by process id.

\begin{figure}[bthp]
    \includegraphics[width=\textwidth]{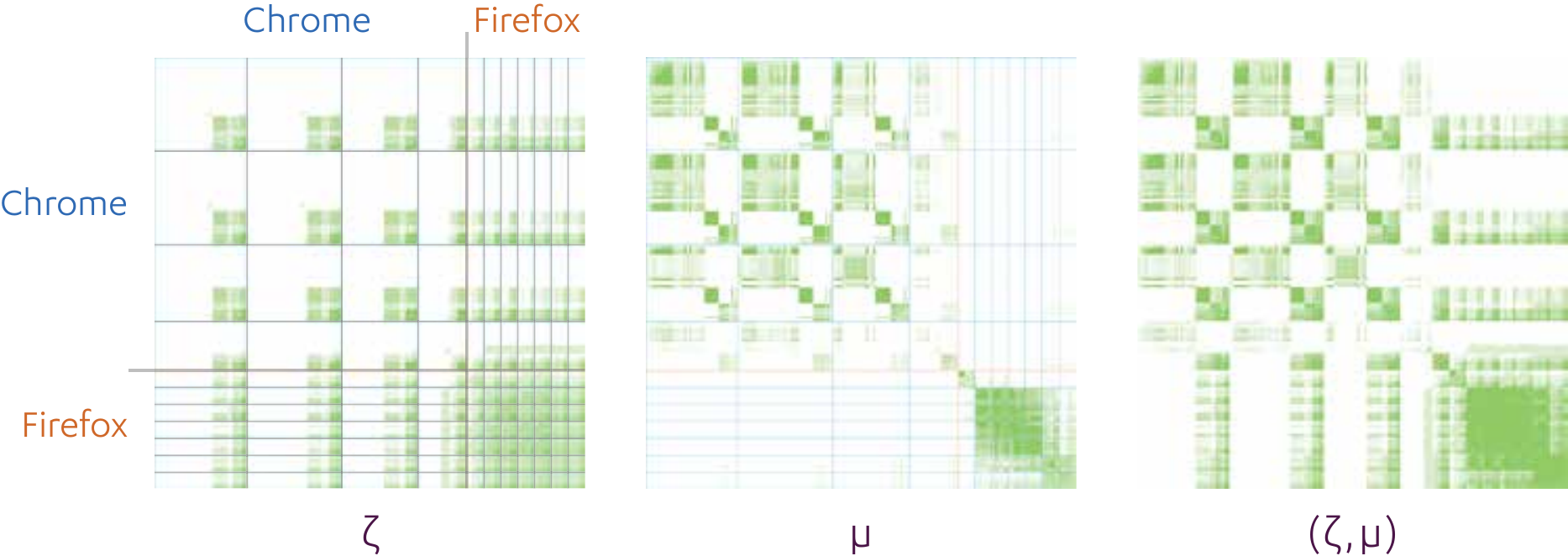}
    \caption{Similarities among the feature vectors collected from executing browsers. Each row and column of the $3073 \times 3073$ matrix is shaded by similarity, darker hues indicating higher similarity.  Feature vectors are gathered by version. The diagonal has been erased.  Left, zeta components are compared; center, mu components; right, the full feature vector.  Chrome and Firefox have similarities in zeta component but little in the mu, while different versions of each have mu similarity without being identical.}
    \label{figMat}
\end{figure}

\autoref{figMat} shows the similarities among the feature vectors created by the different versions of both browsers.  The feature vectors generated by the browser programs were indexed from 1 to 3073, gathered by contiguous ranges of integers for each browser program:  $\brange{c}{12}{1}{659}$, $\brange{c}{17}{660}{1335}$, \ldots , $\brange{f}{44}{2954}{3073}$.  Entry $i,j$ of any matrix in the figure is a similarity between the feature vectors $p_i$ and $p_j$.  The rightmost matrix is the similarity $\zm(p_i,p_j)$, which can be decomposed as the sum of the similarity between its zeta components, $\zm_\zeta(p_i,p_j)$, and its mu components, $\zm_\mu(p_i,p_j)$.  Aside from small corrections due to extreme values, these are essentially the cosine-like distances between the components of the feature vector
\begin{math}
    \zm_\zeta(p_i,p_j) = 
    \theta_\zeta 
    \langle \zeta_i, \zeta_j \rangle
\end{math}
and 
\begin{math}
    \zm_\mu(p_i,p_j) = 
    \theta_\mu 
    \langle \mu_i, \mu_j \rangle
    \,.
\end{math}

\begin{figure}[bp]
    \includegraphics[width=\textwidth]{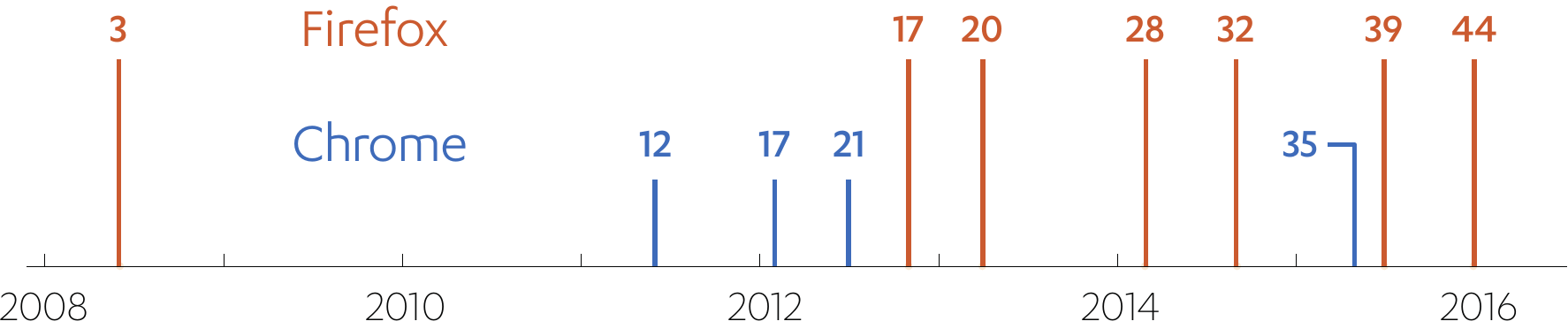}
    \caption{Browser timeline.  Release dates for the versions of the browser programs used to test the sensitivity of the similarity function to changes to the code base.}
    \label{figTime}
\end{figure}

The results in \autoref{figMat} show that the browsers can be compared based just on the average similarity of their feature vectors.  Versions 12, 17, and 21 of Chrome average $64.2$ similarity with each other, whereas their similarity to Chrome version 35 is only $18.9$.  We attribute the large similarity difference to the replacement of the rendering engine from WebKit to Blink, which happened in version 23 \cite{blink}.  The similarities between any Chrome version and any Firefox version are all below $30$.  

The complete set of similarities among the versions of Firefox,
\\ [3ex]
\hspace*{2em}
\begin{tabular}{lrrrrrrr}
 \mbox{} & v3 & v17 & v20 & v28 & v32 & v39 & v44 \\
 \cline{2-8}
 v3 & $126.6$ & $47.4$ & $57.0$ & $42.9$ & $43.8$ & $28.0$ & $25.4$ \\
 v17 & $47.4$ & $272.6$ & $311.7$ & $249.4$ & $176.5$ & $115.2$ & $74.2$ \\
 v20 & $57.0$ & $311.7$ & $405.2$ & $322.0$ & $244.7$ & $139.2$ & $90.8$ \\
 v28 & $42.9$ & $249.4$ & $322.0$ & $330.8$ & $270.8$ & $145.6$ & $87.6$ \\
 v32 & $43.8$ & $176.5$ & $244.7$ & $270.8$ & $361.2$ & $165.9$ & $97.7$ \\
 v39 & $28.0$ & $115.2$ & $139.2$ & $145.6$ & $165.9$ & $323.7$ & $272.2$ \\
 v44 & $25.4$ & $74.2$ & $90.8$ & $87.6$ & $97.7$ & $272.2$ & $294.1$ \\
\end{tabular}
\\ [3ex]
shows the consistency of the similarity and the smooth behavior as the code bases increasingly differ.  While Firefox version 3 differs the most from the other versions, it still has an average similarity of \num{47.4} to version 17.  The similarities slowly decay with increasing version difference (and increasing changes in the code base).

{If we had built a whitelisting system that compared feature vectors with a threshold of $\theta_S = 30$, we would have obtained only one false alert from the browser programs (the one originating from the introduction of Chrome 23). All other browser programs would have been detected as updates. Given the estimated distribution of similarities among different programs, illustrated in \autoref{figDist01}, the whitelisting system would be expected to work with an arbitrary program, as 93\% of the mass of the  \unskip\parfillskip 0pt \par}
\pagebreak
\vspace*{2ex}
\noindent
probability distribution in in \autoref{figDist01}, left, occurs at $\zm(\cdot\, , \cdot) < 30$. That is, a simple comparison of feature vectors with a cutoff of $30$ for the average feature vector similarity can generalize for code development, and only radically new code bases or completely new efforts would produce alerts.

\vspace*{2ex plus 5ex}

\subsection{Activities}

The feature vectors produced by the system are grouped into clusters of high similarity.  These are the tight clusters endowed with the similarity $\zm_{TC}(\cdot\, , \cdot)$ of \autoref{eqTCsim}.  The feature vectors that comprise a tight cluster belong to a sequence of batches collected over a few minutes and need not originate from the same process.  This allows us to detect longer range correlations among the system calls and helps us to identify malware that executes in more than one process.

Each tight cluster corresponds roughly to an activity on the computer. To gain insight into how tight clusters relate to usage, we compared the data generated by two users over the course of 12 weeks. Different programs generated different numbers of tight clusters, reflecting the variety of actions they performed:
\\ [2ex]
\begin{center}
\begin{tabular}{rD{.}{.}{2.0}l}
executable & \multicolumn{1}{c}{clusters} & description \\
\cline{1-3}    
\texttt{chrome} & 66  & Chrome browser  \\
\texttt{WINWORD} & 46  & Microsoft Word \\
\texttt{virtscrl} & 32  &  Lenovo Trackpad support \\
\texttt{spoolsv} & 26  &  Windows print spooler\\
\texttt{SetPoint} & 20  &  Logitech mouse support\\
\texttt{igfxTray} & 7  &  Intel Graphics Accelerator System Tray Helper \\
\end{tabular}
\end{center}
\vskip 1ex
The number of tight clusters contributed by a program decays rapidly with rank. There were 66 tight clusters that contained feature vectors from \texttt{chrome}; the most common clusters were seen \num{373}, \num{146}, and \num{126} times, but \num{43} tight clusters were seen \num{4} times or less.  Most of the tight clusters had several contributing processes (the tight clusters are also not unique to a process).  Some of the tight clusters produced by Chrome are also produced by Internet Explorer.  A preliminary analysis shows that different users tend to generate different distributions of tight clusters when using the same programs.

The compositional nature of programs is reflected in the number of new activities that are created over time.  New activities arise from novel execution paths or user inputs that significantly change the control flow. Limits to programmer productivity and user variability manifest themselves in a human-scale creation rate for new activities.

\begin{figure}
    \includegraphics[height=2in]{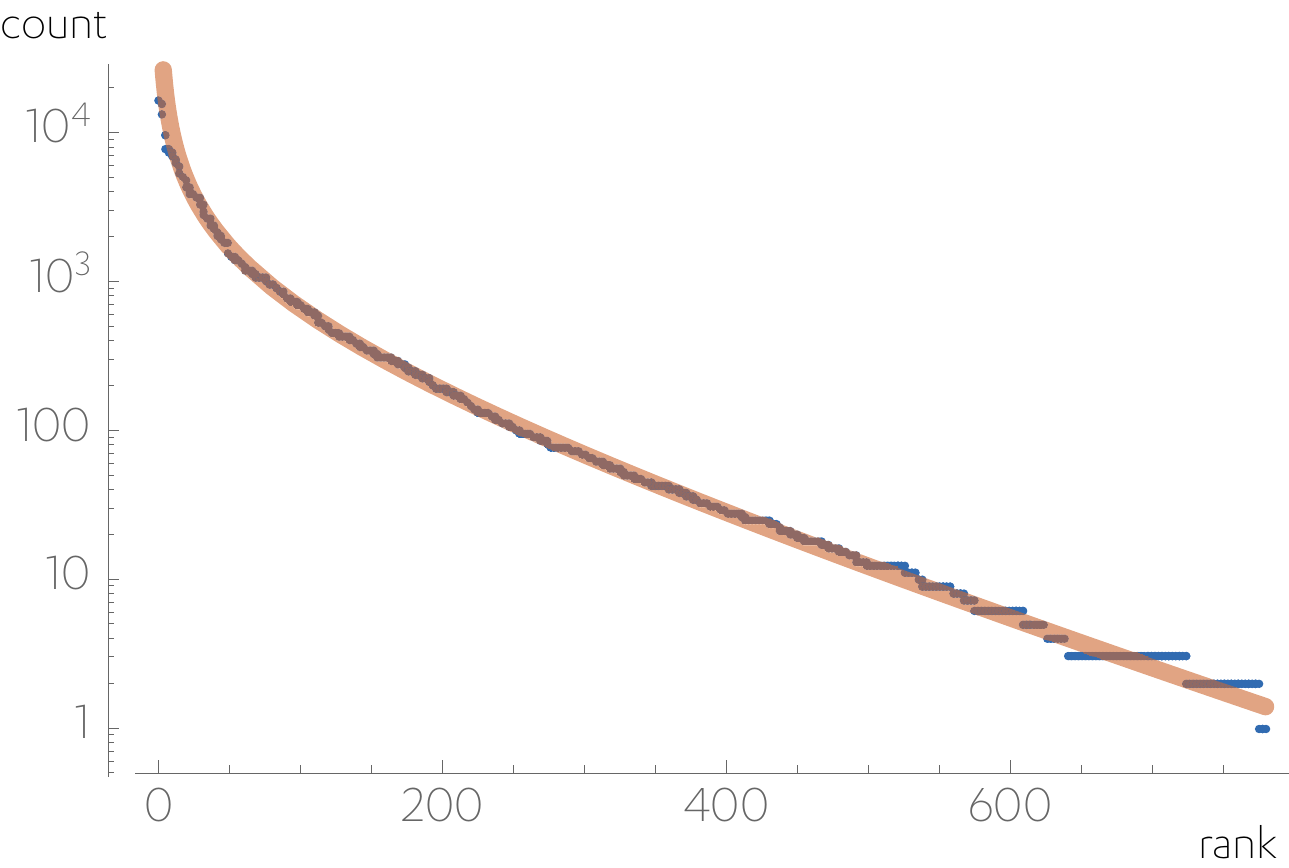}
    \caption{Activity counts sorted by the rank.  The data represent \num{2174} hours of activity.  Among the \num{780} activities seen, the most common one was detected \num{16551} times. The data are well described by a truncated power-law (see the text), suggesting that the number of activities is bounded in practice.}
    \label{figActivities}
\end{figure}

We collected \num{2174} hours of computer activity from a commercial version of our system.  The computers generally were shut down (or hibernated) when users left for the day and on weekends.  Each endpoint contributed a set of tight clusters approximately every 6 minutes (for an average of a few thousand per week).  The set has \num{345463} tight clusters, but only \num{780} distinct ones.  The distribution is shown in \autoref{figActivities} and is well described by $93557 e^{- x/154}(x+5.75)^{-0.916}$.  While distributions with power laws are difficult to fit (see the discussion in \textcite{clauset}), it is simple to verify that a power law function will not reproduce the empirical distribution. (That is, a cutoff, as provided by the exponential, is necessary.)

\begin{figure}
    \includegraphics[height=1.9in]{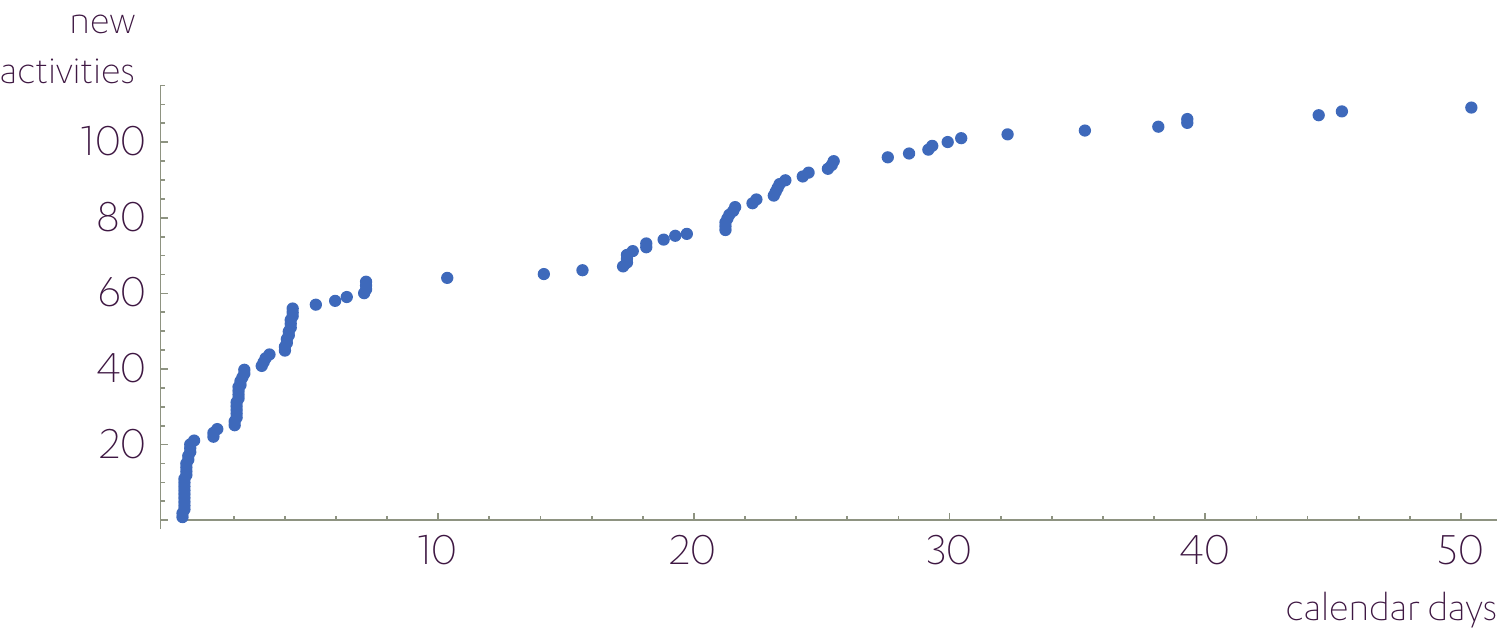}
    \caption{Example rate of new activities for a single endpoint. The rate of new activities continues to slow, with \num{109} new activities observed in \num{54} calendar days.}
    \label{figActRate}
\end{figure}

\autoref{figActRate} shows the number of new activities produced by one computer used by one person over several weeks.  The novelty count rises rapidly when the sensor is first installed, but then slows.  While the rate differs for different classes of users, over a similar time period we have not seen more than \num{160} new activities from a single endpoint.

The new activity creation rate for users (such as the one shown in \autoref{figActRate}) all appear to have a fractional power law behavior where the novelty count grows as $t^\alpha$ for some $\alpha < 1$.  If we interleave several of these processes with different values of $\alpha$ on time scales of minutes, the resulting process will still grow as $t^{\alpha_{c}}$ with a value of $\alpha_{c}$ that is bounded from above by the most productive of the interleaved processes (the one with the largest $\alpha$).  This suggests that the creation rate of new activities can be estimated from the larger dataset obtained by joining streams of data from different endpoints. (We were inspired by the metabook concept of \textcite{metabook}.)

\begin{figure}
    \includegraphics[width=\textwidth]{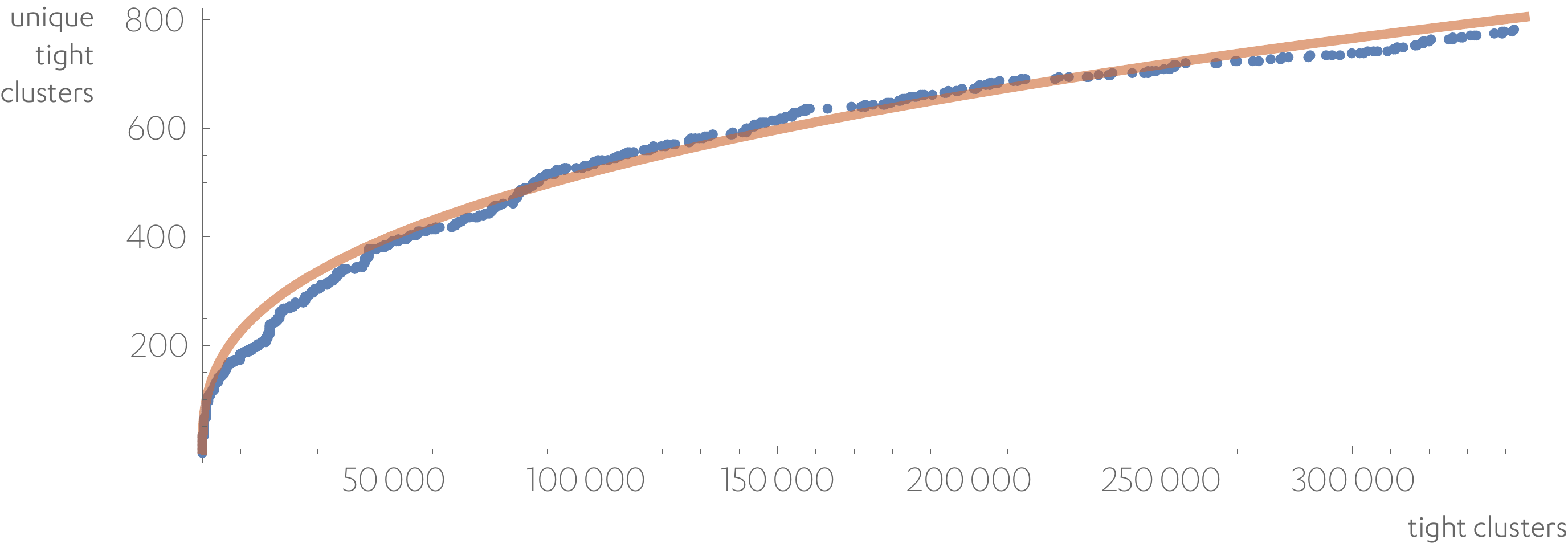}
    \caption{Rate of new activities for all endpoints on a network. As the number of machines active at any one time varied, the number of distinct activities is expressed in terms of the total number of activities observed.}
    \label{figJoined}
\end{figure}

\autoref{figJoined} shows the number of new tight clusters as a function of the total number of tight clusters over the \num{2174} hours of data collected.  The data are well described by a power law with the number of new clusters $m$ given as a function of the total number $N$ of tight clusters collected
\begin{equation}
    % minion math has bad spacings
    m = 8.36 \, N^{\, 0.358}
    \,.
    \label{eqHeaps}
\end{equation}
Larger exponents bound the data from above, and smaller exponents miss earlier data but fit later data better, making (as will be shown) \autoref{eqHeaps} a conservative estimate.

The power law growth of new tight clusters exhibited in \autoref{eqHeaps} is reminiscent of Heaps' description of the number of unique words in a corpus as a function of its size \cite{heaps}.  The relation (known as Heaps' law) can be derived from Zipf's distribution of word frequencies by assuming that the corpus is randomly drawn.  Motivated by the existence of a cutoff in the distribution of tight cluster counts, we consider a modification of Heaps' law when the data are finite.

The overlap of the possibility of an infinite number of different tight clusters with the restriction of an always finite data set makes the problem subtle.  The sequence of tight clusters can be approximated as a generalized P\'{o}lya urn scheme \cite{blackwell} if we make the assumption that each tight cluster is independent of previous ones (which is in our case an approximation).  In this case, the number of tight clusters is expected to be finite and well-characterized by a Dirichlet distribution.  \textcite{zipfTrunc} describe how the Dirichlet is expected to have a parameter vector where the magnitude of its components is rank distributed as an exponentially truncated power law as the number of components grows.  Baek \emph{et al.} also explain that as the size of the collected dataset grows, the number of distinct items seen will grow following an approximate Heaps' law with a parameter $\alpha$ that approaches zero at a rate of $1/\ln N$.

We have checked that our dataset is well described by the results in \textcite{zipfTrunc}.  The probability distribution for groups of tight clusters of size $k$ is described by a distribution proportional to $e^{-k/k_0} x^{-\gamma}$ with $k_0 = 896$ and $\gamma=1.04$.  By a resampling procedure in which only half the data are utilized, we observed that $k_0$ is in the range $[453, 2953]$ and $\gamma$ is in the range $[0.9, 1.1]$.  Parameters outside those intervals yielded visibly poor fits.  We also used a resampling procedure to estimate the exponent $\alpha$ in \autoref{eqHeaps}. Using half the data the exponent was \num{0.41}, suggesting that it decreases with increasing numbers of observations, as expected from finite size considerations.

\subsection{Phylogeny}
\label{secPhylogeny}

To collect the tight clusters of potentially malicious programs, we needed a large and representative sample of executables.  We sampled a number of collections, some containing advanced or specialized malware, and others containing a broad collection. The broad collections were from VX Heavens (covering roughly the years 2005--2010) \cite{vxheavens}, Open Malware (covering roughly 2011--2012) \cite{openmalware}, and Virus Share (covering 2012--present) \cite{vxshare}.  The specialized collections, which contain samples that were found interesting by one or more cybersecurity researchers, were from Contagio Dump / DeepEnd Research \cite{contagio}, malwarechannel \cite{malwarechannel}, theZoo \cite{thezoo}, and KernelMode \cite{kernelmode}. We ran experiments on all PE32 and PE32+ executables sourced from the advanced collections. For the broad collections, we grouped the PE32 and PE32+ samples statically using their import lists, then resampled each group to try to normalize the populations of different malware families.  Small clusters were included whole.

We define a \emph{program family} as the subset of executables that have an orderable set of activities when placed in similar environments.  A program family roughly corresponds to any executable created from the same code base, even if this code base has gone through a small amount of evolution. (Rarely, it could indicate convergent evolution to the level of implementation details.)

All maintained software evolves: bugs get fixed, features get added, and code adapts to changing features of its ecosystem.  Given the nature of a program and its evolution, developers distinguish between releases with a variety of numbering schemes.  In a major.minor release scheme, a program family corresponds to the variations due to patches and minor releases.  The results in \autoref{secBrowserEx} show how several releases of Google Chrome group into one family.

The relationship between a code base and the set of tight clusters that may occur permits a facile construction of a malware phylogeny. We ran each PE32 or PE32+ sample that we had selected repeatedly, to capture as many of the tight clusters as possible. Experiments were run using Windows 7 x64 virtual machines on Xen hosts. Countering all possible anti-virtualization approaches was impossible on the scale of the experiments conducted (for example methods for detecting sandboxes, see \textcite{pafish}).  We avoided easily-detected methods that require an agent on the guests to generate the user events. Instead, we recorded actions and played them back through the vm console (sometimes in a randomized order).

We used our system to extract tight clusters from each experiment. We also constructed a list of tight clusters extracted from identical experiments conducted in the absence of malware. For each sample we pooled all tight clusters not matching any of the tight clusters extracted in experiments conducted in the absence of malware; we then reprocessed the feature vectors from them as though they had originated from a single endpoint. This allowed us to generate a list of tight clusters for each sample. The use of tight clusters allowed us to summarize over $\num{50000}$ hours of activity in the absence of malware in a list of about \num{200} tight clusters. When we used a less extensive list of tight clusters generated from a few hundred hours of usage in the absence of malware, contamination was found in the lists of tight clusters associated with 2\% of the samples. (In production deployments the AI layer handles previously unseen activity; here, it was necessary to mask it so that we could bootstrap the AI layer.)

We then compared all pairs of samples for which we obtained tight clusters, imputing a directed edge $a \rightarrow b$ if each of the tight clusters in $b$ had at least one similar tight cluster in $a$. To avoid both spurious similarities and similarities between unrelated malware that ran inside common Windows processes (such as \texttt{explorer}), we defined tight clusters $a$ and $b$ as similar when $\theta_\mu \langle \mu_a, \mu_b \rangle > 100 $. As there were over 25 billion pairs, it was only possible to conduct the comparison using a reverse index on the $\mu$-words. Even so, it was necessary to conduct the calculation in parts.

The resulting directed graph contains over \num{99} million edges in \num{99235} files. The files were combined by repeatedly: (1) merging two files; (2) merging vertices in loops; and (3) computing the transitive reduction. It was also necessary to conduct this calculation in parts.

The merged and reduced graph was converted to an undirected graph, which contained \num{2.1} million edges and \num{9280} vertices, referencing \num{51648} malicious executables. The graph consists of a few large connected components and a long tail of smaller ones. For each connected component other than the largest we created one family. For the largest connected component we iteratively removed vertices with no in-degree (in the directed graph) until invariant. (These vertices are suprema of fully-ordered subsets of the vertices in the connected component, and where no other apparently identical sample was present. We believe that these samples mainly are downloaders that manifest varied behavior across experiments.) We then created one family for each cluster found in the invariant part using \emph{Mathematica's} \texttt{FindGraphCommunities} with the default algorithm (which is similar to infomap \cite{rosvall2008maps}).

From these steps we obtained \num{1111} families (all of which contain at least two members). The association between observations (here, malicious executables) and families permits us to estimate the population size using statistical procedures similar to those adopted in the ecological literature (see \textcite{efron} for a widely adopted method and \textcite{orlitsky} for a recent one).  In the ecological case, field observations provide a value for the number of specimens of different spices as a function of some sampling effort (typically the observation period).  In the malware case we are interested in estimating the total number of deployed malware families that could execute on a Windows endpoint.  Given the many estimation procedures (see \cite{bunge} for a review) for the total population, reflecting different assumptions about the underlying statistical processes, we applied a variety of these, including Monte Carlo methods.  The estimates for the number of current malware families ranged from \num{1700} to \num{5200}, from which we quote the median of \num{3450} as a reasonable estimate.   Relevant to both the malware detection problem and the functioning and scalability of our AI system, this number is in the thousands and not hundreds of thousands or millions.  With the estimated range for the total number of malware families, one should expect \numrange{1}{2} new malware families appearing per day. 

We then repeated the comparison, graph construction, and clustering processes using those samples that were dissimilar to any other in the test above. In this step we defined tight clusters $a$ and $b$ as similar when $\theta_\mu \langle \mu_a, \mu_b \rangle > 60 $. We refer to families identified in this step as \emph{weak program families}; we found an additional 12 (out of 42 estimated) weak families. Identifying these families is not essential to the operation of the system, but doing so allowed us to either include in a family (or a weak family) or identify as similar to a family (or a weak family) over 99\% of the samples that produced tight clusters in our \num{2.6} million experiments. We refer to the sets of previously detected program families as a \emph{cover}. (In sketching this process we referred to the malware cover as $B$ and the non-malware cover as $G$.)

The estimate of \num{3450} families was obtained using a corpus derived by resampling using static features. The estimator therefore reflects the distribution of malware sampled by the probability of breaching the firewall and antivirus layers (which also rely extensively on static features). This is the relevant population for an endpoint detection and response system; our collection represents an estimated 32\% of the whole. While not strictly required, the existence of a reasonably-sized population with good coverage in the training set provides some assurance that an AI layer will generalize to new and previously-unseen samples.

\subsection{Deep learning}
\label{deepLearning}

For the features we use the zeta component of the original feature vector, a random projection of a subset of the 2-grams present in the subsequences used to create the mu component, and global statistics of both components derived from a partition function over their probabilities (by taking constant size boxes in the method described by Halsey \emph{et al.} \cite{fofalpha}). We trained a 6-layer deep belief network (hidden layer sizes 512, 256, 128, 64, 32) with ReLU activation functions and softmax output on the two-class classification problem $\left( cln, mal \right)$. We used this network to bootstrap class labels for the three-class classification problem $\left( cln, lib, mal \right)$ (where the target label is $cln$ for all non-malware input, $mal$ for malware input classified correctly by the two-class classifier, and $lib$ for malware input classified incorrectly by the two-class classifier). We train an identical (other than the width of the visible layer) network on the bootstrapped class labels and map the softmax output $\sigma(\mathbf{z})$ (where $\mathbf{z}$ is the output of the visible layer) to an odds ratio with
\begin{equation}
    \dfrac{p_{mal}}{p_{cln}} = \dfrac{
        \sigma(\mathbf{z})_{mal} + (1 - \theta_{cln \leftarrow lib}) \sigma(\mathbf{z})_{lib}
    }{
        \sigma(\mathbf{z})_{cln} + \theta_{cln \leftarrow lib} \sigma(\mathbf{z})_{lib}
    }
    \,.
\label{eq3to2}
\end{equation}
Intuitively, mixtures of the \texttt{lib} and \texttt{mal} state correspond to hypothetical malicious programs that can be constructed as \texttt{main.obj} (malicious) and one or more libraries such as \texttt{msvcrt.dll} and \texttt{Crypt32.dll} (not inherently malicious). The bootstrapping step was necessary. With the two-class classifier we obtained $F_1 \approx 0.9$ (without extensive training). With the three-class classifier and $\theta_{cln \leftarrow  lib} = 0.9$ performance increased to $F_1 = 0.995$ (as will be described). We did not further optimize $\theta_{cln \leftarrow lib}$.

For testing we isolated malware that was well-separated by collection time from the training data. The training data was collected prior to March, 2017, the validation data was collected between March and June, 2017, and the test data was collected in July, 2017. We then removed from the test sample all tight clusters similar to any in the training sample. This includes tight clusters that were not present in the cover of \num{1111} families but were obtainable from the training data. We also excluded from the test any samples for which we failed to obtain at least 10 feature vectors (about 80 seconds of execution; this step is necessary to avoid testing on samples that identify the sandbox environment and exit rapidly).

The low rate of new malware families means that estimating the ability of the deep network to generalize requires filtering the test set using a dissimilarity. Otherwise, with many samples, even rare samples in the test set will have with high probability many homologs in the training set that differ only in obfuscation of the static features. As shown by Zhang \emph{et al.}, deep neural networks are capable of memorizing their training set \cite{zhang2016understanding}.

For the non-malware portion of the test tight clusters were randomly sampled from endpoints that were excluded from the training set. Human interface device driver control programs were excluded. (The drivers capture and modify user events, a dangerous behavior; programs in this class should be considered malicious by an AI system. The kernel driver means that signature-based controls are required for these programs anyway.) Among the most recently observed \num{7746} tight clusters there were \num{11} that were considered more likely than not to be potentially malicious: \num{2} of the search indexer; \num{1} of a management utility; and \num{8} of a browser add-on that rewrites web pages before rendering to add reputation information to links. We conservatively considered all of these to be erroneous in calculating the overall performance; excluding the rewriter (which shares a behavior with a broad class of malware) would reduce the false positive rate to \num{0.38} per thousand.

For the malware portion of the test we considered whether a sample is identified as malicious using all of its tight clusters. (The use of all tight clusters is similar to the use of all system calls with more complex network architectures, which may learn to identify thread context switches.) We excluded tight clusters that were not well-characterized (fewer than 10 feature vectors) or not novel (similar to any used to train). We excluded entirely any samples that produced no non-excluded tight clusters.

\begin{table}
\begin{center}
\begin{tabular}{crrrc@{\hskip 0.1in}rlrl}
%& \multicolumn{3}{c}{Confusion} & \multicolumn{4}{c}{Performance} \\
  \multicolumn{4}{c}{confusion}                                                               & & \multicolumn{4}{c}{performance} \\
  \cline{1-4}                                                                                     \cline{6-9}
                           &       & \multicolumn{2}{c}{actual}                               & &    \textsc{fnr}        & $0.0094$ & $F_1$           & $0.9945$ \\
                                     \cline{3-4}
                           &       & \multicolumn{1}{c}{$mal$}   & \multicolumn{1}{c}{$cln$}  & &    \textsc{fpr}        & $0.0014$ & Youden's index  & $0.9892$ \\
\multirow{2}{*}{predicted} & $mal$ & \num{7384}                  & \num{11}                   & &    precision  & $0.9985$ & Error rate      & $0.0053$ \\
                           & $cln$ & \num{70}                    & \num{7735}                 & &    recall     & $0.9906$ &                 &         \\ 
\end{tabular}
\end{center}
\caption{Performance of the neural net layer used to score previously-unseen activities. Results are shown using \autoref{eq3to2} with $\theta_{cln \leftarrow  lib} = 0.9$ and a threshold odds ratio of \num{1} to return the $mal$ class label.}
\label{tblAiLayer}
\end{table}

Results from the deep network should be considered to be a lower bound on performance, as the test samples were not hand-labeled. Few groups have made public the performance of deployable AI systems; our results compare favorably with them ($F_1$ \numrange{0.65}{0.85}, error rate \numrange{0.13}{0.25} \cite{stokes15,thomson_2017}). We also report Youden's index, which better describes the residual work after applying a test (here, the AI). We estimate the residual work with our system to be \num{1.1}\%, with others falling between \num{26}\% and \num{51}\%. The greater than 20-fold reduction in work provides a basis for AI-driven cybersecurity.

%% file: performance.tex
\section{Performance}

We next describe the performance of the system from two perspectives: using traditional machine learning tests to describe the ability of the system to perform a task well and in a generalizable manner; and by a cost analysis to demonstrate that the algorithms and parameters together can be used in a million-endpoint deployment.

\subsection{AI}

In \autoref{tblPerf} we quote the performance of our system.  We use the same corpus and methods used in \autoref{deepLearning} to test the neural net performance (\num{7556} detections from recent malware samples and \num{7746} detections from non-malware; the numbers differ slightly from those in \autoref{tblAiLayer} due to differing exclusion criteria).  For the contamination reasons discussed in \autoref{contamination}, we do not provide performance numbers for the traditional test in machine learning (in which 10\% of the training data is held out). Testing the cover in isolation is difficult because it would require us to produce independently-labeled similar activities.  We can report that in several months of using the system we did not see any false positives for which it was difficult to understand why the system labeled a program as malicious.

One of the reasons to test a machine learning system is to demonstrate that it has not been overfitted and that it generalizes well.  This is difficult to establish if the test data were contaminated with the training data. In a cybersecurity context this may occur if the training and test samples are distinguished only by checksums.  As most malware is obfuscated, it is common to generate many different executables (with different checksums) by randomizing aspects of an executable that do not modify the effects of executing it.  To avoid contamination, many cybersecurity machine learning systems are tested with malware samples that are collected after the training in an effort to include new malware code bases.  This is the approach that we used to report the performance data shown in \autoref{tblPerf}.

\begin{table}
\begin{center}    
\begin{tabular}{llll}
    \multicolumn{1}{c}{system} & 
    \multicolumn{1}{c}{$F_1$} & 
    \multicolumn{1}{c}{error} & 
    \multicolumn{1}{c}{\textsc{fpr}} \\
    \hline  
    Cover + net       & $0.992$ & $0.008$ & $0.0014$ \\
    No homologs       & $0.987$ & $0.010$ & $0.0014$ \\
    Net only          & $0.995$ & $0.005$ & $0.0014$
\end{tabular}
\end{center} 
\caption{Performance of the system under different conditions. The line “Cover + net” best reflects the expected performance under deployed conditions. For details see the text.}
\label{tblPerf}    
\end{table}

The performance shown for the “Cover + net” system was measured by comparing the test detections to all tight clusters in the cover, then feeding those that did not match well to the deep neural net described in \autoref{deepLearning}.  This is the common “by-date” test used to test machine learning systems for cybersecurity.  

Even with the by-date approach the test data may still be contaminated. Samples in the test set may arise from existing malicious codebases either from code reuse or because the malware was only recently detected and added to the corpus. To mitigate that possibility, we adopted an approach that uses the phylogeny we established in \autoref{secPhylogeny} to eliminate any activities that match an existing activity in the cover. The cover allows us to perform a close approximation of an attack with never before used executables.  By eliminating any tight clusters generated from malware samples that match the cover, we eliminate from the test set most of the malware that is a slight modification of previously-seen malware (the \emph{homologs}).  The results are shown on the “No homologs” line; we see very little loss in performance.

\enlargethispage{\baselineskip}
We reproduce the results of using only the AI layer (reported in \autoref{deepLearning}) on the line “Net only”. The slightly increased performance reflects the fact that the exclusion process described in \autoref{deepLearning} preferentially eliminates failure cases for the network. These cases (which match activity from malware experiments but were not included in the cover) may reflect non-malicious behavior early in the malware execution pathway (frequently for sandbox evasion). The network prefers to label them $cln$ despite many training examples with the $mal$ and $lib$ class labels.

\subsection{Scaling}

In our system data streams from different endpoints only interact with each other through the shared database of observed activities. Activities are timestamped with their initial observation time and can be shared among a fleet of servers; if an activity is detected nearly simultaneously on two or more servers, then agreeing to keep only the activity with the earliest detection time among similar activities creates an eventually consistent collection. To model how the system scales to very large networks, we need only consider the server-side compute cost, the number of false positives (which creates human costs, as best practices in computer security suggest they be investigated), and the implications of a larger working set of activities that have been recently observed. 

The number of activities will grow slowly.  Given the theoretical robustness of the distributions we have observed, we expect that a fractional power law function (\autoref{eqHeaps}) will bound the number of unique tight clusters as a function of the number of observed clusters, even when the number of observed clusters is much larger than in our current datasets.  Larger datasets will have even smaller exponents, improving the performance of the system.  Even with the exponent of \num{0.358}, one million endpoints should produce about \num{21} thousand new activities in one week, which can be handled by an in-memory database.  We can ignore the cost of eventual consistency among databases; even with one billion endpoints and neglecting improvements to scale, \autoref{eqHeaps} estimates the new activity rate to average fewer than 1 per second.

The interaction between the sensor and the compute server occurs through a REST API \cite{fielding}, which allows the system to leverage existing hyper-scalable web technologies. Using CPUs to compute and discarding data after processing (which allows us to dispense with some database costs), we observe server side compute costs under 1 watt per continuously protected endpoint. (Using a sampling strategy and correcting for the duty cycle, this would allow a single 10 MW data center to protect tens to hundreds of millions of endpoints. Bandwidth usage would not saturate modern interconnects.)

% at a 10% sampling rate and accounting for duty cycle we would
% need 96 gigabits actual into the building
% this is way less than the petabit within-region interconnect, see
% https://azure.microsoft.com/en-us/blog/how-microsoft-builds-its-fast-and-reliable-global-network/?cdn=disable

% 672000000*0.1*(50/168)*650*8/1024^3

The false positive rate is determined by the performance of the AI layer and the rate of new activities (as false positives, once investigated, can be made not to recur). The AI layer described in \autoref{deepLearning} has a false positive rate of \num{1.3} per \num{1000}. With one million endpoints and assuming an exponent of \num{0.358}, the system would average \num{26.2} false positives per week. (Further scaling improvements would be observed on larger networks.)

The size of the working set has implications for detecting malware, as the likelihood of a spurious match between a new activity and one in the database grows with the database size. (To characterize users by the distribution of the observed activities, it suffices to substitute a probability distribution for the one-hot vector that encodes the activity.) Estimating the number of false negatives requires an estimate of the infection rate, but if 1\% of the endpoints were executing malware each in a single process, with one million endpoints the system would average \num{62} false negatives per week from working set size effects.

%% file: discussion.tex
\section{Discussion}
\label{secDisc}

We have described an AI system that produces a stream of scored descriptors that reflect the activities of an endpoint.  The system has an $F_1$ score of 99.1\% at a false positive rate of 0.14\%.  This is achieved by using a similarity function for the activities that generalizes well and that groups the bulk of the malware seen over the last 10 years into \num{1111} program families.  A version of the system has been deployed.  We have argued that it could scale with the observed (or better) characteristics to a network of one million endpoints.  We now place these claims into the context of previous work and accepted practices in the cybersecurity and machine learning communities.

\subsection{Similarity function}

The performance of the system is a reflection of the properties of the similarity function $\zm(p_a, p_b)$ (\autoref{eqSim}).  The similarity changes smoothly with modifications to a code base, as the examples of Chrome and Firefox show. Between 2011 and 2016 the loc for each program more than tripled, yet the similarity function identifies similarities between the most distantly related versions shown in \autoref{figTime}. The similarity function is at the appropriate scale for following a code base that only changes as fast as programmers can modify it.

We can use the similarity function to define a program family as the set of executables that have strong $\mu$-similarity, as we did in creating the malware phylogeny (\autoref{secPhylogeny}).  Because the system characterizes in-execution malware, it is not hampered by static file obfuscations and steganographic techniques; because it does not rely on emulation techniques, it is not hampered by flow-control hiding. These properties allow a better characterization of existing malware.  We estimated in \autoref{secPhylogeny} that the total number of malware families was \num{3450}, much smaller than generally believed.

The malware phylogeny is constructed using the similarity function $\zm_{TC}(\cdot\,,\cdot)$.  Several algorithms in machine learning do not have an explicit similarity function, but if they are consistent (converge with large training samples), then they are also local in the feature space (values of the function depend only on nearby points), as discovered by \textcite{locality}.  Those algorithms could, in principle, reproduce our construction of the malware phylogeny, estimate the number of families, and remove homologs for testing AI subsystems, as described in \autoref{contamination}.

The cover-matching part of our AI is a $k$-nearest neighbor algorithm, which is often eschewed in machine learning because of the difficulty in finding nearest neighbors \cite[][see sections 13.3--5]{hastie}. In our case the feature vectors can be indexed using information retrieval techniques because the mu part is a very sparse vector of token counts.

\subsection{Malware in the wild}
\label{contamination}

Reported numbers on malware vary dramatically depending on how they are classified.  Based on file checksums, VirusTotal receives over \num{50000} new executables every day that are not labeled by any of the scanners that it hosts \cite{vt}.  Detection rates increase with a lag, but some of those executables are tagged as generic malware rather than as a member of any of the families that antivirus vendors track \cite{lastline,avclass}.  Such numbers cannot reflect different program families in the sense of different code bases. A Fermi estimate \cite{fermi,purcell} for the number of groups developing malware suggests not more than a few hundred groups engaged in malware development: a number of states sponsoring multiple groups and a handful of additional black market enterprises. If each of these groups releases two substantially new malicious programs a year, we should expect between one and two new (not detected by existing engines) pieces of malware daily in VirusTotal (and maybe \num{4000} over 10 years). We attribute the difference to the poor detection capabilities of current systems. (Using ssdeep \cite{kornblum2006identifying} to cluster samples from Virus Share, \textcite{tsukasaoi} achieved about a 17-fold reduction, which is still over \num{1000}-fold too high.)

If we accept that there is much less distinct malware than what is estimated from the number of distinct checksums, then it is clear that testing of malware detection AI systems needs to be modified.  One of the reasons that machine learning systems are tested is to ascertain their ability to generalize.  If a randomly-chosen tenth of the data is used for testing, it is very likely that copies of the same malware (but with different checksums) will be in both the training and the test sets.  For the common case in which the labeled corpus consists of $n$ malware samples and $n$ benign programs with a tenth used for testing, a malware that occurs $r$ times in the set is $1-e^{-r/10} + O(n^{-1} r^2 e^{-r/10}/40)$ likely to be in the training and test sets.  To a first approximation the probability is independent of the corpus size, $2n$.  In a corpus with \num{250000} pieces of malware, \num{46} copies of the same malware (code, not hash) would be sufficient to have a 99\% probability of having one of the 46 in both the training and test sets.  With just two copies the probability is 18\%.

\subsection{Performance comparisons}

The system was designed to be deployable (the technology has been incorporated into a commercial product), which limits the set of algorithms that can be used, bounds the needed performance, and caps operational costs per endpoint. The stream of scored activities can be combined with other data to create a system that comes to a decision and performs an action.  We have concentrated on the case of an alert that is used to trigger an investigation into possible exploitation of an endpoint (how we quoted the performance). The data have also proven useful for the insider threat aspect of cybersecurity; that use case requires a time series of actions that were performed on the endpoint.

\begin{table}
\begin{center}    
\begin{tabular}{lllll}
    \multicolumn{1}{c}{system} & \multicolumn{1}{c}{$F_1$} & 
    \multicolumn{1}{c}{error} & 
    \multicolumn{1}{c}{\textsc{fpr}} & \multicolumn{1}{l}{reference} \\
    \hline  
    Cover + net  & $0.992$ & $0.008$ & $0.0014$ & [this paper] \\
    Android LSTM & $0.915$ & $0.084$ & $0.0710$ & \cite[Ref.][]{androidlstm} \\
    ESN + LR     & $0.857$ & $0.125$ & $0.0014$ & \cite[Ref.][]{stokes15} \\
    SOC          & $0.686$ & $0.455$ & $0.9000$ & [see text]
\end{tabular}
\end{center} 
\caption{Comparison of various AI approaches to security operations. A composite estimate of the expected performance of current SOC practices is included for reference.}
\label{tblOther}    
\end{table}

For comparison we have included a few other results from the literature, shown in \autoref{tblOther}.  Production workflows use analysts in security operations centers (SOCs) to interpret the results of \emph{ad hoc} collections of tools that process log data, network data, and (in some cases) sparse subsets of system calls.  We are not aware of any published tests of the performance of such workflows, but public comments from practitioners suggest high false positive rates (\textsc{fpr}).  As a baseline guide we provide an estimate using 0.9 as the \textsc{fpr}, 0.01 as the false negative rate (\textsc{fnr}), and a test in which an equal number of malware and non-malware programs had to be identified.  Its performance is indicated by the line marked “SOC” in \autoref{tblOther}, with an $F_1$ score of $0.686$ and an error rate of $0.455$ (defined as the number of erroneous decisions as a fraction of the total).  The only other result we identified that quotes a performance with a \textsc{fpr} of under 1\% was the work of Pascanu \emph{et al.} \cite{stokes15}; in their work ROC curves present data starting at a \textsc{fpr} of \num{1e-4}. From those curves we computed the numbers for the line “ESN + LR”. At an \textsc{fpr} of 1\%, the same system achieves an $F_1$ score of $0.952$.  Another approach that appears well suited to identifying malware with system calls employs LSTM networks.  We quote the results of \textcite{androidlstm} (which do not go below a \textsc{fpr} of 7.1\%, unsuitable for production work) on the “Android LSTM” line in the table.

\subsection{Sensor}

The sensor collects the complete stream of system calls using facilities provided by the operating system.  We use the ETW (Event Tracing for Windows) framework.  ETW has a great advantage for professional (as opposed to research) software; it is developed and maintained by the authors of the operating system. Other systems that attempt to characterize the functioning of a computer by examining the system calls use some form of function hooking (in which the system calls are made to execute code that provides new functionality, such as recording that the function was called).  Hooking breaks one or more software invariants (locking strategies, calling conventions, timing constraints, and many other hard-to-enumerate properties) that programmers rely upon and performant and compliant software demands.   Hooking also slows down the computer; factors of 2 are common \cite{Deng:2012ee}.  The slowdown arises from the loss of cached context when the operating system switches between kernel and user modes. 

To keep the load placed by the sensor on the CPU below 1\%, the arguments to the system calls are not collected.  Intuition and previous work \cite{canali2012quantitative} suggest that better performance can be achieved by using the arguments of the system calls.  By concentrating on only the calls, we were forced to develop detection techniques that reflect the actions of processes rather than the target of those actions (a verbs versus nouns distinction).   Battery-powered devices connected over cellular networks are limited in the amount of power that they can allocate to data transmission; this gives preference to systems with low bandwidth requirements.  The sensor averaging \num{5} seconds executing and \num{10} seconds idling will upload an average of \num{650} \si{bytes/s}.

\subsection{Strategies}

Cyber attackers use automated means that render defenses that incorporate manual steps infeasible at scale. Van Dijk \emph{et al.}~\cite{flipit} captured the essence of the problem by framing it as a resource utilization contest; alerting and triage are among the least efficient steps in the defender's kill chain \cite{kumar2017practical}. AI could reverse this asymmetry, but only if it is sufficiently performant to operate autonomously. Our system has several characteristics that make this possible. The cover (\autoref{secPhylogeny}) will often detect previously seen malware with just one batch of data. The false negative rate of $0.0094$ (\autoref{deepLearning}) permits fully automated procedures for all but the largest or most secure organizations. The ability to detect executables that are variations of a code base (\autoref{secBrowserEx}) will force attackers to use increasingly expensive strategies in an effort to maintain access.